\documentclass{article}

     \usepackage[final, nonatbib]{tccml_neurips_2020}

\usepackage[utf8]{inputenc} 
\usepackage[T1]{fontenc}    
\usepackage{hyperref}       
\usepackage{url}            
\usepackage{booktabs}       
\usepackage{amsfonts}       
\usepackage{nicefrac}       
\usepackage{microtype}      
\usepackage{graphicx}
\usepackage{amsmath}
\usepackage{amssymb}
\usepackage{booktabs}
\usepackage{multirow}
\usepackage[dvipsnames]{xcolor}
\usepackage[american]{babel}

\def\RR{{\mathbb R}}    
 
\def\EE{{\mathbb E}}    
\def\11{{\mathbf 1}}    

                                 \def\b1{{\mathbf 1}}

               \def\cD{{\mathcal D}}            \def\cL{{\mathcal L}}     

\def\SSIM{\operatorname{SSIM}}

\title{Predicting Landsat Reflectance with Deep Generative Fusion}

\author{%
  Shahine Bouabid\thanks{Work completed as part of the Cervest Residency program. Correspondence to \texttt{jev@cervest.earth}} 
  \quad Maxim Chernetskiy \quad Maxime Rischard \quad Jevgenij Gamper \\
   Cervest Ltd. \\
   London, UK \\
}

\begin{document}

\maketitle

\begin{abstract}
Public satellite missions are commonly bound to a trade-off between spatial and temporal resolution as no single sensor provides fine-grained acquisitions with frequent coverage. This hinders their potential to assist vegetation monitoring or humanitarian actions, which require detecting rapid and detailed terrestrial surface changes. In this work, we probe the potential of deep generative models to produce high-resolution optical imagery by fusing products with different spatial and temporal characteristics. We introduce a dataset of co-registered Moderate Resolution Imaging Spectroradiometer (MODIS) and Landsat surface reflectance time series and demonstrate the ability of our generative model to blend coarse daily reflectance information into low-paced finer acquisitions. We benchmark our proposed model against state-of-the-art reflectance fusion algorithms.
\end{abstract}

\section{Introduction}

Amid climate-induced stress on land resource management, detecting land cover changes associated with vegetation dynamics or natural disaster in a timely manner can critically benefit agriculture, humanitarian response, and Earth science. Orbiting remote sensing devices stand as prime assets in this context, continuously providing multi-spectral terrestrial surface imaging~\cite{van2013fieldcopter}. They assist in decision-making with valuable large-scale perspectives that enable assessment of vegetation indices~\cite{crop_yield_modis, phenology_california}, forest wildfire spread~\cite{review_fire_ecology}, and flooding risks~\cite{flood_management, flood_inundation_modelling}. 

Unfortunately, a trade-off between spatial and temporal resolution in this type of imaging is often observed. For example, sensors with a large scanning swath are able to cover wide regions at once, enabling high revisit frequency but also resulting in poor spatial resolution. The Moderate Resolution Imaging Spetroradiometer (MODIS) sensors~\cite{700993, 701081} offer a daily revisit cycle that is well suited to tracking rapid surface changes but only capture ground resolution cells ranging from 250-500m. Conversely, sensors with greater spatial resolution provide more precise views but take longer to revisit. For instance, Landsat missions~\cite{landsat, ROY2014154} provide imagery at 30m resolution, which is sufficient to discern individual crops. However, these missions suffer from 16-day revisit cycles and frequently encounter issues of cloud occluding. These technical constraints undermine the availability of free remote sensing imagery with sufficient coverage to meet the needs of precision agriculture or geohumanitarian actions.

\emph{Surface reflectance fusion} is the task of combining imagery products with different characteristics to synthesize a mixed reflectance product with sufficient spatiotemporal resolution for land-cover monitoring. The complementarity of low-paced fine-resolution Landsat acquisitions and coarse daily MODIS updates makes them natural candidates for surface reflectance fusion. Statistical approaches have been successfully introduced~\cite{STARFM, ESTARFM, USTARFM} that leverage consistency between Landsat and MODIS in reflectance~\cite{landsat_modis_consistentcy}, but these techniques rely heavily on the temporal density of the available satellite acquisitions. Alternatively, deep learning has shown promise when applied to specific remote sensing tasks~\cite{fuentes2019sar, grohnfeldt, wang, zotov2019conditional, deepsum, rudner2019multi3net}. Notably, deep generative models have demonstrated their capacity to fuse multiple frames with low spatial resolution into a higher resolution one~\cite{deudon2019highres, deepsum}.

In this work, we propose to address surface reflectance fusion with deep generative models applied to Landsat and MODIS images. Our contribution is threefold: {\bf (i)} we propose a deep generative framework to estimate Landsat-like reflectance; {\bf (ii)} we introduce a dataset of paired Landsat and MODIS reflectance time series; {\bf (iii)} we conduct a quantitative and qualitative evaluation against state-of-the-art reflectance fusion algorithms.

\section{Background}

The introduced dataset contains time series of co-registered Landsat and MODIS reflectance images at various sites. Let $\cD = \{\mathbf{l}_{t_i}, \mathbf{m}_{t_i}\}_{i=1}^N$ denote a time series with $\mathbf{l}_{t_i}, \mathbf{m}_{t_i} \in \RR^{B \times W \times H}$ representing the Landsat and MODIS fields, respectively, at date $t_i$. The image dimensions $B, W, H$ are respectively the number of spectral bands, image width, and image height. Note that the MODIS frames have been georeferenced and resampled to Landsat resolution. Furthermore, Landsat and MODIS spectral bands typically cover the same domains such that they can be considered consistent in reflectance~\cite{landsat_modis_consistentcy}. 

Consider a coupled pair of Landsat and MODIS images at date $t$, $\{\mathbf{l}_t, \mathbf{m}_t\}$, and a corresponding pair at some future prediction date $t_p$, $\{\mathbf{l}_{t_p}, \mathbf{m}_{t_p}\}$. The surface reflectance fusion problem may then be framed as predicting $\mathbf{l}_{t_p}$ given $\{\mathbf{l}_t, \mathbf{m}_t\}$ and $\mathbf{m}_{t_p}$. Formally, this entails that homogeneous pixels in spatial position $[i, j]$ should satisfy $\mathbf{l}_{t_p}[:,i,j] = \mathbf{m}_{t_p}[:,i,j] + \mathbf{l}_t[:,i,j] - \mathbf{m}_t[:,i,j]$. 

\subsection*{Adaptative Reflectance Fusion Models}

Gao, et al.~\cite{STARFM} propose a spatial and temporal adaptative reflectance fusion model (STARFM) based on the above relationship to estimate daily Landsat-like surface reflectance. By extending the interpolation range to a surrounding spatial, temporal and spectral window around pixel $[i, j]$, they better account for heterogeneous pixels and changes between acquisition dates.

STARFM has since been successfully applied to estimate daily Landsat-like reflectance and assist vegetation monitoring~\cite{STARFM_NDVI, STARFM_dryland, STARFM_vegetation_monitoring, STARFM_dense_ts_generation}. Improved version such as Enhanced STARFM (ESTARFM)~\cite{ESTARFM} and Unmixing STARFM (USTARFM)~\cite{USTARFM} further focus on the treatment of endmembers from heterogeneous coarse pixels and allowed to better cope with fine grained landscapes and cloud contamination.

\subsection*{Deep Generative Imagery in Remote Sensing}

Neural networks have displayed compelling performance at generative tasks in remote sensing such as cloud removal from optical imagery~\cite{grohnfeldt}, SAR-to-Optical image translation~\cite{fuentes2019sar, wang, zotov2019conditional} or spatial resolution enhancement~\cite{deudon2019highres, lanaras, deepsum}. In particular, work on multi-frame super-resolution has demonstrated the capacity of deep neural networks to combine information from multiple low resolution images to produce finer grained outputs~\cite{deudon2019highres, deepsum}. Rudner, et al.~\cite{rudner2019multi3net} have also pointed out how fusing satellite products with complementary spatial, temporal and spectral information can benefit tasks such as flooded building segmentation. 

Generative adversarial networks (GANs)~\cite{NIPS2014_5423} have demonstrated impressive capacity for natural image generation~\cite{brock2018large}, image translation tasks~\cite{pix2pix2017} and single-image super-resolution~\cite{berthelot2020creating}. These successes highlight the ability of GANs to impute learnt details from images with poor resolution. The conditional GANs (cGANs) paradigm has received particular interest in remote sensing as it allows generated samples to be conditioned on specific inputs. For example, conditioning on radar images has shown promising performance at generating realistic cloud-free optical remote sensing imagery~\cite{grohnfeldt, wang, fuentes2019sar}.

\section{Method}

In the following, we address the task of predicting a Landsat-like image $\mathbf{l}_{t_i}$ at date $t_i$ given MODIS image at the same date $\mathbf{m}_{t_i}$ and last known Landsat image $\mathbf{l}_{t_{i-1}}$. Although we use Landsat and MODIS reflectance, we bring to the reader's attention that this fusing rationale remains compatible with other remote sensing products.

\subsection*{Supervised cGAN for Landsat Reflectance Prediction}

Conditional GANs are a class of generative models where a generator $G$ learns a mapping from random noise $\mathbf{z}$ and a conditioning input $\mathbf{c}$, to an output $\mathbf{y}$. That is, $\mathbf{y} = G(\mathbf{z}| \mathbf{c})$. The generator is trained adversarially against  a discriminator $D$, which in turn learns to estimate the likelihood $D(\mathbf{y}| \mathbf{c})$ of sample $\mathbf{y}$ being real or generated by $G$.

Suppose we want to estimate the Landsat surface reflectance $\mathbf{l}_{t_i}$ at date $t_i$. Doing so would require ground-level structural information of the site, which could be obtained from the last known Landsat image $\mathbf{l}_{t_{i-1}}$. However, information about actual reflectance at the current date $t_i$ could be inferred from the corresponding coarse resolution MODIS image $\mathbf{m}_{t_i}$. Let $\mathbf{c}_{t_i}$ denote the concatenation of these two fields along spectral bands, $\mathbf{c}_{t_i} = \operatorname{Concat}\left(\mathbf{l}_{t_{i-1}}, \mathbf{m}_{t_i}\right)$.

We approach this problem using the cGANs framework and propose to train a generator to predict Landsat-like surface reflectance, $\mathbf{\hat l}_{t_i} = G(\mathbf{z}| \mathbf{c}_{t_i})$. We augment the objective with a supervised component to constrain $G$ to output images close to $\mathbf{l}_{t_i}$. Specifically, we incorporate an $L_1$ penalty to capture low-frequency image components, while inducing less blurring than an $L_2$ norm. Also, a structural similarity index measure (SSIM)~\cite{SSIM} criterion fosters the generation of higher frequency components expressed through local pixels dependencies.

Subsequently, the generative reflectance fusion problem is written as the minimax two-player game

\begin{equation}
\min_G\max_D \,\cL_{\text{cGAN}}(G, D) + \alpha \cL_{L_1}(G) + \beta \cL_{\SSIM}(G) ,
\end{equation}

where the loss terms are given by $\cL_{\text{cGAN}}(G, D) = \EE\left[\log D(\mathbf{l}_{t_i}|\mathbf{c}_{t_i})\right] + \EE\left[\log\left(1 - D(G(\mathbf{z}|\mathbf{c}_{t_i}\right)|\mathbf{c}_{t_i}))\right]$, $\cL_{L_1}(G) = \EE\left[\|\mathbf{l}_{t_i} - G(\mathbf{z}|\mathbf{c}_{t_i})\|_1\right]$ and $\cL_{\SSIM}(G) = \EE\left[1 - \SSIM(\mathbf{l}_{t_i}, G(\mathbf{z}|\mathbf{c}_{t_i}))\right]$. Expectations are taken over all possible $\{\mathbf{l}_{t_i}, \mathbf{c}_{t_i}\}$ pairs and $\alpha, \beta > 0$ are supervision weight hyperparameters.

\subsection*{Dataset}

The study area is in the department of Marne, within the Grand Est region of France. It is mostly constituted of crops, forest and urban areas. We acquire Landsat-8 30m reflectance imagery for 14 dates spanning from 2013 to 2020, and MODIS 500m surface reflectance product (MCD43) for the same dates. Images are reprojected to the same coordinate system and MODIS frames are bilinearly resampled to the Landsat resolution and bounds. We limit the spectral domain to red, near-infrared, blue and green bands. Contaminated image regions are discarded using quality assessment maps.

We extract 256$\times$256 non-overlapping registered Landsat and MODIS patches at each date, resulting in 548 distinct patches and a total of 5671 Landsat-MODIS pairs when accounting for multiple time steps. The dataset is split into 383, 82 and 83 patches locations for training, validation and testing.

\section{Experiments}

\subsection*{Experimental setup}

Inspired by the success of \texttt{pix2pix}~\cite{pix2pix2017} based approaches in remote sensing~\cite{grohnfeldt, wang, fuentes2019sar}, we use U-Net architecture~\cite{unet} for the generator and a PatchGAN discriminator~\cite{pix2pix2017}. U-Net skip connections directly pass the spatial structure from the last know Landsat frame across the network. This allows the intermediate layers to only learn variations from this baseline. We provide stochasticity in the training process using dropout layers. The PatchGAN discriminator jointly classifies local regions of the image, unlike image-level discriminators that process the entire input as a whole. This localized analysis fosters generation of high-frequency components to mimic realistic image textures. All code is made available~\footnote{\href{https://github.com/Cervest/ds-generative-reflectance-fusion}{\texttt{https://github.com/Cervest/ds-generative-reflectance-fusion}}}.

Quantitative evaluation of generated samples relies on several full-reference image quality metrics: peak-signal-to-noise ratio (PSNR), SSIM~\cite{SSIM}, and spectral angle mapper (SAM)~\cite{SAM} scores.

\subsection*{Results}

\begin{figure}[t]
\label{figure:predictions}
\centering
\includegraphics[width=\linewidth]{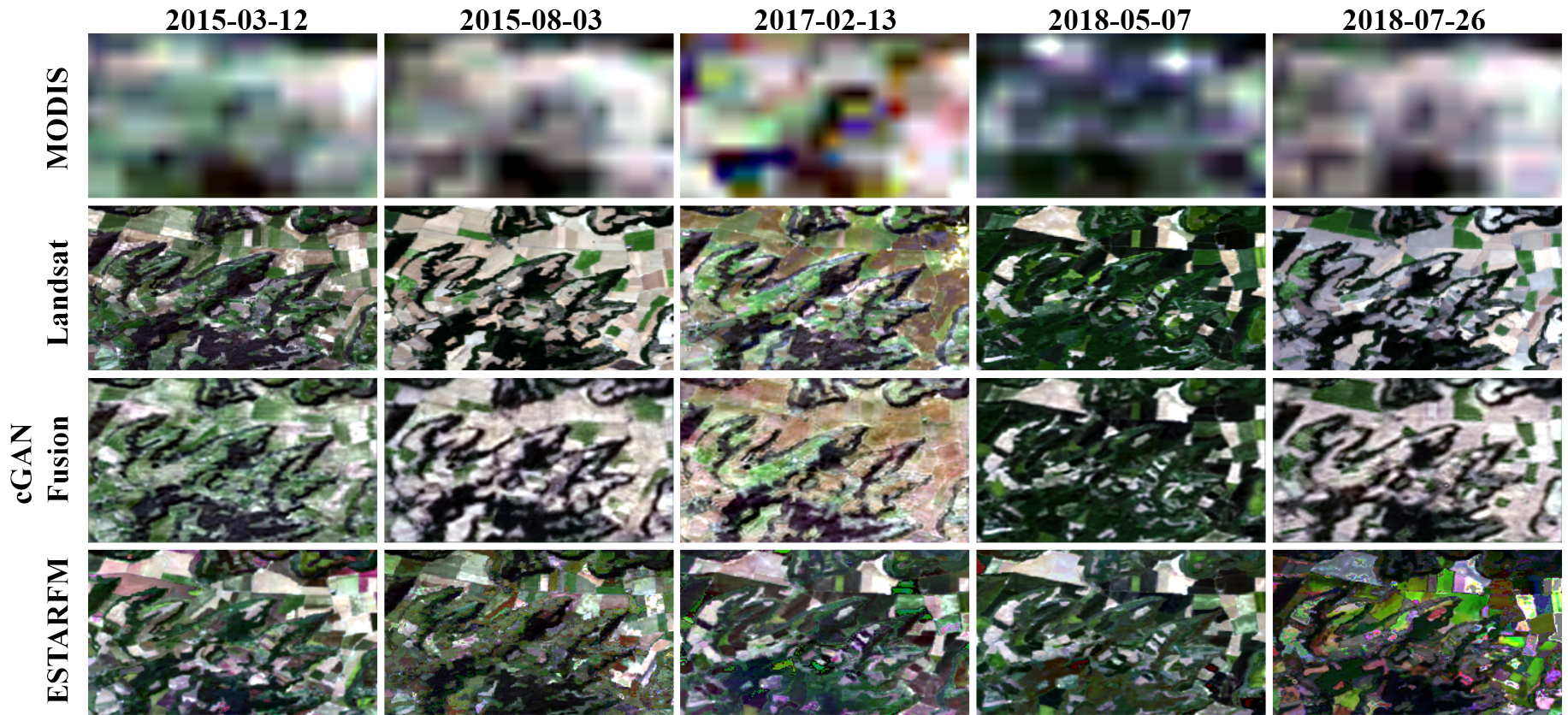}
\caption{Example of Landsat-like surface reflectance prediction time series, each column is a different time step; "cGAN Fusion" is infered from MODIS at same date and previous Landsat; "ESTARFM" is infered from MODIS at same date and Landsat-MODIS pairs at previous and next dates}
\end{figure}

\vspace*{-0.2cm}
\begin{table}[h]
\centering
\footnotesize
\begin{tabular}{lccccccccc}\toprule
Method                                       & \multicolumn{4}{c}{PSNR} & \multicolumn{4}{c}{SSIM} & SAM (10\textsuperscript{-2}) \\\midrule
Band                                         & NIR    & R   & G   & B   & NIR    & R   & G   & B   &     \\\midrule
Bilinear Upsampling                          &  20.0  & 19.0    &  21.0   &  21.1   &  0.568      &  0.550   &  0.633   & 0.639    &  3.87   \\
ESTARFM~\cite{ESTARFM}                       &  19.6  & 20.2& 21.8& 22.3& 0.555  &  0.640   &  0.688   &  0.696   &  4.88   \\
cGAN Fusion + $L_1$                          &  22.1   &   21.8  &  23.7   &  23.8   &   0.675     & 0.697    &  0.747   &  0.747   &  2.75   \\
cGAN Fusion + $L_1$ + SSIM  & {\bf 22.3}   & {\bf 22.0}   & {\bf 23.9}    & {\bf 24.0}    &  {\bf 0.694}      & {\bf 0.714}    & {\bf 0.761}    &  {\bf 0.760}   & {\bf 2.70}   \\\bottomrule
\end{tabular}
\caption{Image quality scores on testing set; cGAN models scores are averaged over 3 independently trained models}
\label{table:IQ_scores}
\end{table}
\vspace*{-0.3cm}

We compare our method against ESTARFM implementation~\cite{cuESTARFM}, which requires at least two pairs of Landsat-MODIS images at distinct dates and MODIS reflectance at prediction date. Table~\ref{table:IQ_scores} highlights the substantial improvement in image quality metrics on an independent testing set for which the SSIM objective has not been optimized.

Figure~\ref{figure:predictions} provides a qualitative comparison of generated Landsat-like surface reflectance from the cGANs approach and the ESTARFM method. We see that the cGANs approach demonstrates flexibility at capturing and blending MODIS reflectance into Landsat images, sensibly respecting the shape of ground-level instances. ESTARFM, being more conservative with regards to image sharpness, produces quite realistic looking samples. However, it struggles to recover the correct spectral values when images used as inputs are too temporally distant from the prediction date.

Furthermore, we observe that the SSIM loss term improves the stability of the adversarial training process and prevents GANs-induced up-sampling artifacts such as checkerboard patterns~\cite{odena2016deconvolution}. It encourages the discriminator to instill realistic contrast levels and image structures (e.g., fields boundaries and forested/urban areas), which have finer details. However, we notice a subsequent forcing in luminance that pushes generated crops toward lighter tones, while an $L_1$-only supervision renders more faithful colors.

\vspace*{-0.2cm}
\section{Conclusion}

Using cGANs, we can develop implicit generative models capable of producing visually promising results at surface reflectance fusion. We demonstrate their capacity to faithfully capture the broad features from coarse reflectance images and fuse it into detailed images. This can help circumvent limited access to public imagery with sufficient spatial and temporal resolution to assist precision agriculture and humanitarian response.

\section{Acknowledgments}

We would like to dedicate a special mention to Andrew Glaws for his support and insightful comments on this work as part of the Tackling Climate Change with Machine Learning workshop at NeurIPS 2020
Mentorship Program.

{\bibliographystyle{ieee_fullname}
\bibliography{neurips_2020}
}

\newpage
\appendix

\section{Experimental and Implementation details}

The U-Net architecture used as generator in all experiments uses 5 down and upsampling levels, leading to 6$\times$6$\times$1024 latent features. We use batch normalization and parameterized ReLU activations on encoding and decoding layers. Models details are provided in Table~\ref{table:architectures}.

The PatchGAN model used as a discriminator, similarly downsamples input into a 15$\times$15$\times$1 feature map on top of which a sigmoid activation is employed to classify the underlying region of each cell. Batch normalization and leaky ReLU activations are used on the downsampling layers. See Table~\ref{table:architectures} for architecture details.

The Adam optimizer~\cite{Adam} with safe initial learning rate of 3$\cdot$10\textsuperscript{-4} --- decayed by 0.99 at each epoch --- is used for training both generator and discriminator. To make up for the additional guidance provided to the generator by the supervision objectives, we backpropagate on the discriminator twice as often. A batch size of 64 is used during training. Finally, in order for the different components of the objective function to land in comparable ranges, we use supervision weights $\alpha=0.1$ and $\beta=100$.

\newgeometry{left=1cm,right=1cm}
\begin{table}
    \footnotesize
    \centering
    \begin{tabular}{l}\toprule
         {\bf \normalsize U-Net}\\\toprule
         {\bf Encoder} \\\toprule
         Conv2D(in=8, out=64, kernel=4, stride=2, padding=1)\\\midrule
         Conv2D(in=64, out=128, kernel=4, stride=2, padding=1)\\
         BatchNorm(channels=128)\\
         PReLU()\\\midrule
         Conv2D(in=128, out=256, kernel=4, stride=2, padding=1)\\
         BatchNorm(channels=256)\\
         PReLU()\\\midrule
         Conv2D(in=256, out=512, kernel=4, stride=2, padding=1)\\
         BatchNorm(channels=512)\\
         PReLU()\\\midrule
         Conv2D(in=512, out=1024, kernel=4, stride=2, padding=1)\\
         BatchNorm(channels=1024)\\
         PReLU()\\\midrule
         Conv2D(in=1024, out=1024, kernel=4, stride=1, padding=1)\\
         BatchNorm(channels=1024)\\
         PReLU()\\
         \toprule
         {\bf Decoder} \\\toprule
         ConvTranspose2D(in=1024, out=1024, kernel=4, stride=1, padding=1)\\
         BatchNorm(channels=1024)\\
         Dropout(p=0.4)\\
         PReLU()\\\midrule
         ConvTranspose2D(in=2048, out=512, kernel=4, stride=2, padding=1)\\
         BatchNorm(channels=512)\\
         Dropout(p=0.4)\\
         PReLU()\\\midrule
         ConvTranspose2D(in=1024, out=256, kernel=4, stride=2, padding=1)\\
         BatchNorm(channels=256)\\
         PReLU()\\\midrule
         ConvTranspose2D(in=512, out=128, kernel=4, stride=2, padding=1)\\
         BatchNorm(channels=128)\\
         PReLU()\\\midrule
         ConvTranspose2D(in=256, out=64, kernel=4, stride=2, padding=1)\\
         BatchNorm(channels=1024)\\
         PReLU()\\\midrule
         ConvTranspose2D(in=128, out=64, kernel=4, stride=2, padding=1)\\\toprule
         {\bf Output Layer}\\\toprule
         ConvTranspose2D(in=64, out=4, kernel=3, stride=1, padding=1)\\\bottomrule
    \end{tabular}
    \quad
    \begin{tabular}{l}\toprule
         {\bf \normalsize PatchGAN } \\\toprule
         Conv2D(in=12, out=128, kernel=4, stride=2, padding=1)\\
         LeakyReLU($\alpha$=0.2) \\\midrule
         Conv2D(in=128, out=256, kernel=4, stride=2, padding=1)\\
         BatchNorm(channels=256)\\
         LeakyReLU($\alpha$=0.2)\\\midrule
         Conv2D(in=256, out=512, kernel=4, stride=2, padding=1)\\
         BatchNorm(channels=512)\\
         LeakyReLU($\alpha$=0.2)\\\midrule
         Conv2D(in=512, out=512, kernel=4, stride=2, padding=1)\\
         BatchNorm(channels=512)\\
         LeakyReLU($\alpha$=0.2)\\\midrule
         Conv2D(in=512, out=1, kernel=4, stride=1, padding=1)\\
         Sigmoid()\\\bottomrule
    \end{tabular}
    \caption{U-Net Generator and PatchGAN Discriminator architectures}
    \label{table:architectures}
\end{table}
\restoregeometry

\section{Figures}

\begin{figure}[h]
\label{figure:scatterplot}
\centering
\includegraphics[width=0.8\linewidth]{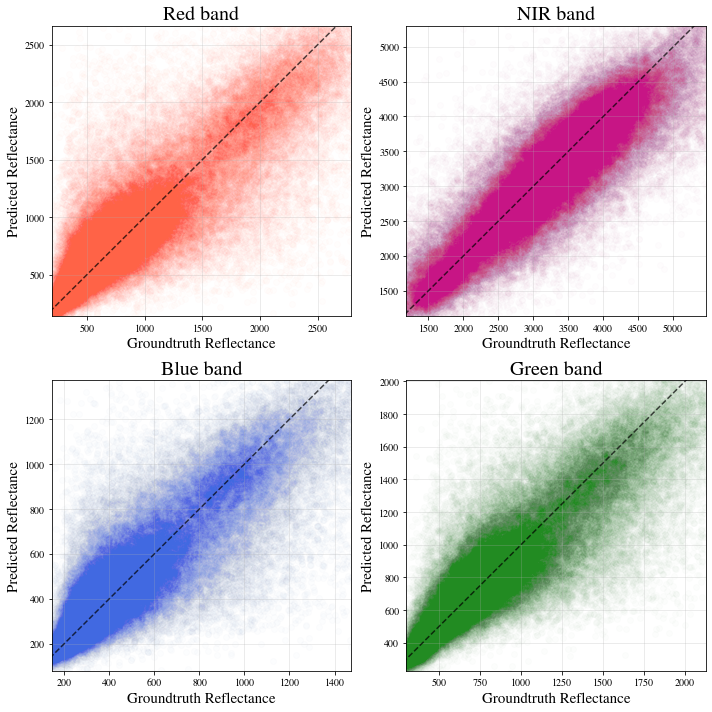}
\caption{Scatter plot of predicted pixel values by band against groundtruth Landsat pixel values on the testing set}
\end{figure}

\end{document}